\input harvmac
\skip0=\baselineskip
\divide\skip0 by 2
\def\tmpsp{\the\skip0}

\def\skipthis#1{{}}

\def\p{\partial}
\def\ph{\phi}
\def\Ph{\Phi}

\Title{\vbox{\baselineskip12pt\hbox{hep-th/9911104}
\hbox{HUTP-99/A062}}}{Modeling a network of brane worlds}

\centerline{Soonkeon Nam\foot{Permanent Address :
Dept. of Physics, Kyung Hee University; Seoul, 130-701, Korea, nam@string.kyunghee.ac.kr}
} 
\bigskip\centerline{Department of Physics}
\centerline{Harvard University}
\centerline{Cambridge, MA 02138}
\centerline{\it nam@pauli.harvard.edu}

\vskip .3in \centerline{\bf Abstract} 
We study junctions of supersymmetric domain walls in $N=1$ supergravity 
theories in four dimensions, coupled to a chiral superfield with
quartic superpotential having $Z_3$ symmetry.
After deriving a BPS equation of the domain wall junction, we
consider a stable hexagonal configuration of network of brane junctions,
which are only approximately locally BPS.
We propose a model for a mechanism of supersymmetry breaking without loss of stability,
where a messenger for the SUSY breaking comes from the neighboring anti-BPS junction world,
propagating along the domain walls connection them.

\smallskip
\Date{11/99}
\lref\rubakov{V. Rubakov and M.E. Shaposhnikov, Phys. Lett. {\bf 125B} (1983) 136.}
\lref\antoniadis{I. Antoniadis, Phys. Lett. {\bf 246B} (1990) 377.}
\lref\add{N. Arkani-Hamed, S, Dimopoulos, and G. Dvali, Phys.\ Lett.\ {\bf B429} (1998) 35.}
\lref\addk{N. Arkani-Hamed, S. Dimopoulos, G. Dvali, and N. Kaloper,
hep-th/9907209.}
\lref\rs{L. Randall and R. Sundrum, hep-th/9905221, hep-th/9906064.}
\lref\st{K. Skenderis and P.K. Townsend, hep-th/9909070. }
\lref\gog{M. Gogberashvili, hep-ph/9904383.}
\lref\brandhuber{A. Brandhuber and K. Sfetsos, hep-th/9908116.}
\lref\cs{C. Csaki and Y. Shirman, hep-th/9908186.}
\lref\nelson{A.E. Nelson, hep-th/9909001.}
\lref\dfgk{O. DeWolfe, D.Z. Freedman, S.S. Gubser, and A. Karch,
hep-th/9909134.}
\lref\bc{K. Behrndt and M. Cvetic, hep-th/9909058.}
\lref\bs{I. Bakas and K. Sfetsos, hep-th/9909041.}
\lref\bpsjunc{G.W. Gibbons and P.K. Townsend, Phys.\ Rev.\
Lett.\ , {\bf 83} (1999) 1727.}
\lref\bpsdom{K. Skenderis and P.K. Townsend, hep-th/9909070.}
\lref\Boucher{W. Boucher, Nucl.\ Phys.\ {\bf B242} (1984) 282.}
\lref\Townsend{P.K. Townsend, Phys.\ Lett.\ {\bf 148B} (1984) 55.}
\lref\lps{H. Lu, C.N. Pope, and E. Sezgin, Phys.\ Lett.\ {\bf B371} (1996)
46-50.}
\lref\cg{A. Chamblin and G.W. Gibbons, hep-th/9909130.}
\lref\saffin{P.M. Saffin, to appear, Phys. Rev. Lett., hep-th/9907066.}
\lref\mpp{E.W. Mirabelli, M. Perelstein, and M.E. Peskin, Phys. Rev. Lett. 
{\bf 82} (1999) 2236.}
\lref\mp{E.W. Mirabelli and M.E. Peskin, Phys. Rev. {\bf D58} (1998) 065002.}
\lref\cnamone{J. Casahorran and Soonkeon Nam, Int. J. Mod. Phys. {\bf A6} (1991) 5467.}
\lref\cgr{M. Cvetic, S. Griffies, and S. Rey, Nucl. Phys. {\bf B381} (1992) 301.}
\lref\rps{B.S. Ryden, W.H. Press, and D.N. Spergel, Ap. J.
{\bf 357} (1990) 293.}
\lref\cht{S.M. Carroll, S. Hellerman, and M. Trodden, hep-th/9905217.}
\lref\oins{H. Oda, K. Ito, M. Naganuma, and N. Sakai, hep-th/9910095.}
\lref\susybreak{N. Arkani-Hamed and S. Dimopoulos, hep-ph/9811353.}
\lref\vilenkin{M. Aryal, A.E. Everett, A. Vilenkin, and  T. Vachaspati,
Phys. Rev. {\bf D34} (1986) 434.}
\lref\cv{S. Cecotti and C. Vafa, Commun. Math. Phys. {\bf 158} (1993) 569.}

For many years it was believed that the extra dimensions, as are required from
string/M-theory, are compactified with sizes of order of $10^{-33}$cm.
The ensuing disparity of  the electroweak scale of $~10^3$GeV and the Planck energy
scale of $~10^{18}$ GeV made impossible to directly probe the regimes of quantum
gravity.

Meanwhile an alternative idea of our matter made of zero modes
trapped on a topological defect (3+1)dimensional, embedded
in a higher dimensional universe was also proposed\rubakov\antoniadis.
Recent proposals of large extra dimensions in a similar philosophy
has provided exciting new possibilities of addressing long standing problems such as cosmological
constant problem, hierachy problem and supersymmetry breaking\add .
Explicit realization of the idea was presented in Refs.\gog\rs.
The basic tenet of these works is that the standard model resides on a 3 dimensional brane or
intersection of branes in the higher dimensional spacetimes where as the gravity is progating
throughout entire dimensions.
For the simplest model of thin branes intersecting, it was shown that the
gravity indeed localizes on the intersection\addk\cs\nelson.

It would be satisfactory if we could model this brane (or
intersection/juction of branes) from a higher dimensional theory, where the
brane is a stable solution of a nonlinear equation, having origin in
supergravity or string/M-theory.
Models of smooth solution has been considered in Ref.\dfgk\st.
Here they considered five dimensional gravity coupled to a single scalar
to model a stable domain wall, without supersymmetry.
However, further efforts to obtain the smooth solutions directly from $D=5$
$N=2$ supergravity theories runs into problems such as curvature singularity on
the location of the brane\bc. $D=5$, $N=8$
gauged supergravity theories were also considered\bs.
Also a distribution of D3 branes were considered\brandhuber.
The constraints to these models of brane world should also come
from phenomenology point of view.

The model we will be considering in this letter will be that of domain wall junction.
Just as a BPS domain wall breaks half of the supersymmetry, there are BPS domain wall
junction configurations which break one quarter of supersymmetry\bpsjunc\cht.
In these configurations, stability is directly linked to the supersymmetry of the system.
So, one phenomenological question which arises in this context is how to break supersymmetry, without
losing the stability of the configuration.
Here, we propose a way of achieving both, stability as well as breaking of supersymmetry.
For this purpose, we consider a network of junctions. 
Actually there is a work on infinite network of junctions of domain walls 
which are only locally approximately BPS states\saffin. Such a network enjoys
stability against local fluctuations, as well as the domain wall thickness
does not exceed the size of the domains bound by them. However, it has been
been done in the context of rigid supersymmetry. So in order to apply to the
`brane-world' senario, we first have to establish similar BPS objects within
the context of supergravity.

Let us consider the simplest model which has a single flat 3-brane
embedded discontinuosly in the ambient geometry\add.
\eqn\action{S=\int d^D x\sqrt{-g} \left[  {1\over 2} R -{1\over 2}(\p \ph)^2 -V(\ph)
\right],}
in the mostly positive metric.
One is interested in the special form of the potential \st
\eqn\pot{
V(\Phi)=2(d-2)^2\left({\p W \over \p \Phi}\cdot{\p W \over \p \Phi} -
{d-1\over d-2} W^2  \right),}
which is inspired by supergravity\cnamone.

The advantange of the form is that the spacetime is now specified by the 
solution of the nonlinear equation for the dilaton, in the absence of the
gravity\cg. We then can integrate the second equation for the metric.
At the critical points of the superpotential, the potential is negative, 
and we do have anti-de Sitter space.
The stability equations of small fluctuations take the form of
supersymmetric quantum mechanics and are devoid of harmful tachyonic modes.

We would now like to explore the possiblity of having a similar smooth solution
for the case of intersecting branes or junctions of them.
One of the motivation for this is to have nontrivial tension on the junction
itself, which might affect the gravitational field nearby.
(However, it will not affect the asymptotic geometry because, the junction energy will
be source of gravity in one higher dimensional space than the branes.)
 Another is to have a stepping stone model to a more string theoretic
explanation of the brane configuration.

In order to have intersections/junctions which allow a four dimensional
Minkowski spacetime at the overlapping region, we need to have at least two
large extra dimensions and two real scalars.
The simplest would be the junction of three 4 branes in six dimensional
embedding space.
The strategy is thus to consider a six dimensional gravity coupled to two
real scalars with some potential, which has at least three minima,
which allows brane junctions.
Analogous to the domain wall case of five dimensions, we want to find the form
of the potential which makes the BPS equation of the scalar fields satisfy
BPS equation for junction in the absence of gravity\cg.
\eqn\decouple{
\Phi' = {\p W\over \p \Phi}, A' = -{1\over d-2} |W|.}
We can calculate the metric by quadrature, and at the vacua, when the 
derivative of the superpotential vanishes, we have automatically 
AdS space in the asymptotic region.
So we first have to  study BPS equations for intersecting domain walls,
in the absence of gravity\bpsjunc.
For this we have to generalize the system to that of complex scalar and 
two extra dimensions.

As a toy model we consider the $1+1$ dimensional brane world. This can be
achieved by considering $N=1,D=4$ supergravity theories which allow 
domain wall solutions\cgr.
For that purpose, consider an $N=1$ $D=4$ globally supersymmetric field theory
of a chiral superfield $\Ph$ with a complex scalar field
$\ph$ and a fermion $\chi$, which arises as an effective field theory.
The bosonic part of the lagrangian is in terms of the K\"ahler potential
$K( \Ph , \bar\Ph )$, and a holomorphic superpotential
$W( \Ph )$:
\eqn\lagfourdim{L = \eta^{ab}K_{\ph\bar \ph} \p_a \ph\p_b \bar \ph-
K^{\ph\bar \ph}|\p_\ph W(\ph)|^2.}
This model was studied in Ref.\bpsjunc\cht\ and was shown to admit
brane junction solutions, when there are more than three minima of the superpotential.
These solutions admit one quarter of the
supersymmetry, with a single Hermitian supercharge.
The issue of junctions and supercharge was also studied in 
Ref.\oins.

The anticommutators of two left handed supercharges has central charges as follows:
\eqn\susyalgebra{\{Q_\alpha, Q_\beta\} = 2i(\sigma^k\bar\sigma^0 )_
\alpha^\gamma \epsilon_{\gamma\beta} Z_k.}
The anticommutators between left- and right-handed supercharges has a 
contribution from other supercharges $Y_k$, $k=1,2,3$,

\eqn\sysalgebratwo{\{Q_\alpha, \bar Q_{\dot\alpha} \} =
2(\sigma^\mu_{\alpha{\dot\alpha}} P_\mu +\sigma^k_{\alpha{\dot\alpha} }Y_k ).}
where $P_\mu$ are the energy momentum four vector.
$Z_k$ are complex and $Y_k$ are real.
If we have single domain wall which is nontrivial only in one dimension,
then $Y_k$ vanishes and $Z_k$ is non-vanishing.
However, when the field configuration at infinity is nontrivial in two
dimensions, as in the cases of domain wall junctions, $Y_k$ can be
nonvanishing.
This is actually proportional to the area in field space spanned by the
fields of the solution, as measured by the K\"ahler metric, and is the
junction mass\cht, and can have negative values also\oins.

Now let us consider an $N=1$ locally supersymmetric theory, whose bosonic
part is given as follows:
\eqn\laggravfourdim{
e^{-1}L = -{1\over 2}R+ g^{\mu\nu}K_{\ph\bar \ph} \p_\mu \ph\p_\nu \bar \ph-
e^{K}(K^{\ph\bar \ph}|D_\ph W(\ph)|^2-3|W|^2).}
where $e = |{\rm det} g_{\mu\nu}|^{1/2}$, and $D_\ph W = e^{-1} (\p_\ph e^K W )$.
The convention we are adopting are those of Ref.\cgr, where $\gamma ^\mu = e^\mu_a\gamma^a$ where
$\gamma^a$ are the flat spacetime Dirac matrices satisfying $\{\gamma^a, \gamma^b\} = 2\eta^{ab}.$
$a=0,1,2,3$. We also have put $8\pi G =1$.
We use the Weyl basis for the gamma matrices.
The projection operators are $P_{R,L} = {1\over 2}(1\pm
i \gamma^5)$, and  $K^{\phi\bar\phi} = 1/\p_\phi\p_{\bar\phi} K$, and
$K_{,\phi} = \p_{\phi} K$.

Existence of domain wall for the simplest case of $Z_2$ symmetric 
case was demonstrated in Ref.\cgr.
\eqn\Ztwo{K= \ph\bar \ph,\ \ \  W = {1\over 3} \ph^3 - a \ph.}
For the junctions to exist, we need at least three extrema of the superpotential,
so we consider the following:
\eqn\Ztwo{K= \ph\bar \ph,\ \ \ W = {1\over 4} \ph^4 - b \ph.}
In the global case one has three isolated minima at
$\ph_n= b^{1/3}e^{2\pi n i /3}$, $n = 0,1,2$. In Ref.\cgr, it was argued that domain walls cannot
exist in such a theory, because all the geodesics in the space of the superpotential
has to go through the origin, which is not a vacuum. However it turns out that we can have
BPS domain wall junctions which has two spatial coordinate
dependences rather than one as in the cases of domain walls of Ref.\cgr.

To see this we first obtain the BPS equation for the scalar field.
This comes from the vanishing of supersymmetry transformation of fermion of the chiral superfield.
 The spin 1/2 field $\chi$ transforms as
\eqn\susytrans{\del_\epsilon \chi =
-\sqrt{2} e^{K/2}K^{\phi\bar\phi}(D_{\bar\phi}\bar W P_R +
 D_{\phi}W P_L)\epsilon -i\sqrt{2}(\p_\nu \phi P_R + \p_\nu\bar \ph P_L)
\gamma^\nu \epsilon,  }
and the gravitini transform as
\eqn\gravitinitrans{
\delta_ \epsilon \psi_\mu
= \left[2(\p_\mu +{1 \over 2} \omega^{ab}_\mu \sigma_{ab})
+i e^{K/2} ( W P_R + \bar W P_L)\gamma_\mu
-{\rm Im}(K_{,\phi}\p_\mu \phi)\gamma^5\right] \epsilon.}

Since we are interested in constructing static brane junction solution, which in the thin limit
gives us the patches of AdS spaces\addk,
the ansatz for the vierbein we choose is $e^a_\mu ={\rm diag}(A^{1/2}(x,z),A^{1/2}(x,z),
A^{1/2}(x,z),A^{1/2}(x,z))$, which has just a conformal factor, for the spacetime metric:
\eqn\metric{ds^2 = A^2(\vec{z}) \eta_{\mu\nu} dx^\mu dx^\nu.}

In order to satisfy the BPS equation one has to have
$\epsilon = (\epsilon_1,  i\zeta\epsilon_1, \zeta\epsilon_1, -i\zeta\epsilon_1)$
where $\zeta=\pm 1$. We will call the solutions of $\zeta =1$ case a BPS configuration and
$\zeta=-1$ case an anti-BPS configuration.

The BPS equations we find are
\eqn\BPSEQ{\eqalign{
& \delta_\epsilon\chi = 0 : (\p_x +i\p_z)\ph = -\sqrt{A}e^{K/2}K^{\ph\bar\ph}
D_{\bar\ph}\bar W \cr
& \delta_\epsilon\psi_x = 0  : (\p_x+i\p_z)\log A = -2 \sqrt{A} e^{K/2}\bar W ,\cr
& \delta_\epsilon\psi_y = 0 : \p_z \log A= -2{\rm Im}(K_{,\phi}\p_x \phi)+2ie^{K/2}\bar W
\sqrt{A}, \cr
& \delta_\epsilon\psi_z = 0 : (\p_x+i\p_z)\log A = -2 \sqrt{A} e^{K/2} \bar W, \cr
& \delta_\epsilon\psi_t = 0 : \p_x \log A = 2{\rm Im}(K_{,\phi}\p_z \phi)-2e^{K/2}\bar W
\sqrt{A}.\cr} }
Just as in the case of the domain walls, these equations satisfy Einstein's equation,
and in the thin limit we have three domain wall junctions with angles of
$2\pi/3$ fixed. Here we no longer have the constraint on the geodesic on $W$
space, and wall junctions exist. We see that the scalar field will be solved
as a function of $x+iz$. Unlike the rigid superymmetric case, here we have
nontrivial coupling to the gravity through $\sqrt{A}$ and it is in general
difficult to solve it analytically. However, it inherits the general structure
of BPS equations of rigid supersymmetric case\bpsjunc.

One can imagine that there can be other types of superpotentials which
admit domain junction, such as $W = \ph ^5/5 - \phi^2/2$, which has four
extrema, three at the edges of an equilateral triangle and one at
$\phi=0$. In $1+1$ dimensional case, it was shown that the kinks interpolating
the edges of the triangle does not exist\cv, based on the existence of
an intersection number of two cycles around the extrema. It would be
interesting to have similar arguments for the higher dimensional cases.

Note that unlike the domain wall case, the relative phases of the components
of the spinor are completely fixed, signaling further breakdown of
supersymmetry. However, the fact that there is a consistent BPS equation
available shows that there is still some supersymmetry available, which is one
quarter of the original supersymmetry.
As mentioned earlier, there will be a contribution to the central charge of supersymmetry algebra,
which will be interpreted as the junction energy.
The contribution from the junction energy will not matter for the asymptotic geometry, because it will
a source of gravity for one higher dimension than the domain walls thus falls off faster.
However, for detailed study of evolution of the universe or other phenomenological studies,
it might be of use.

The BPS configuration of junction of three domain walls exists even in the context of supergravity.
It will be straightforward to generalize this to higher dimensions.
Brane world on a junction can be viewed as a model with two types of extra
dimensions. One is the directions transverse to all the domain walls and has
truely the nature of the bulk. The other is along the direction of the domain
walls. In the model we considered there is one such direction. This direction
is different from the transverse directions to the domain walls because,
massless modes can still be captured here. So we need not have only gravity
mode along this direction. In fact, these massless modes can propagate along
the domain wall in the speed of light.
Now, on the junction, the massless modes can be further trapped because the
domain walls get thinner away from the junction. However, if we have another
junction within a finite distance, joined by one of the domain walls, then the
massless modes can travel over to the other junction with finite probability,
because the wall thickness will grow as we approach another junction.
Of course this would be difficult to see in a model with thin
branes, but within the smooth model we have considered above, this is
certainly the case. Of course an explicit demonstration of this would require
a numerical study. All in all, {\it there can be information exchange between
the junctions, if they are joined together by a domain wall.}

Now that we have a stable BPS junction, we come back to the issue of
breaking supersymmetry. Attempts of having a supersymmetry breaking mechanism
was proposed  in Ref.\mp\susybreak, which involve bulk messengers, either
from a SUSY broken hidden sector, or a supersymmetric source of a massive bulk
messenger. It is not clear whether the stability of the domain wall is not
at stake with the spontaneous breaking of supersymmetry.
A way of stablizing the configuration is to consider a network of brane
junctions, forming an array. Given that a stable static triple intersection
exists, one can also think about a networks of domain walls. Certain 
configuration of network
is meta-stable, and for the case of $Z_3$ model we have studied, BPS
junctions have to be joined with anti-BPS junctions\saffin. For the case of
$Z_3$ invariant theory, hexagonal domains, like graphite, was shown to be
stable under local fluctuations.
Numerical simulation showed that it is
stable, as long as the domain sizes are greater than the thickness of the
walls. We can have metastable configuration of hexagonal structures. These are
non-BPS (but almost BPS) configurations. Maybe some of the effect of the
neighboring anti-BPS junction give rise to the breakdown of supersymmetry at
our junction. If we have finite thickness of the domain wall, some of the
massless modes can propagate along the walls. So the presence of of the
neighboring anti-BPS junctions will be known to us on a BPS junction, by some
messenger\mp\susybreak. So this should give rise to a way of having
supersymmetry broken, without losing the stability against perturbations.
However, there is some difference between the mechanism here and the ideas of
Ref.\mp, where one needs to have a spontaneously broken supersymmetry in a
hidden sector residing on the  other brane, at a finite distance apart along
the extra dimension. The physics on the other brane world is much different
from ours. Here we do quite similar brane world. This model can be regarded as
a toy model of the mechanism proposed in Ref.\susybreak. To be more specific,
on our junction world, one particular combination of original supercharges,
say, $Q_1$ is left unbroken, and the states will be invariant under the action
of the charge. On the other hand, the states originating from the neighboring
anti-BPS junction world will be invariant by the action of different linear
combination, say $Q_2$. Then they will certainly be not invariant under $Q_1$
and will be seen as a messenger of supersymmetry breaking.
Of course, if the array forms a regular lattice, then there will be Bloch
wave functions along the lattice.

Numerical simulations of the hexagonal lattice shows that as long as the
domain sizes are greater than the width of the walls.
The parameter which controls the supersymetry breaking will be related to the ratio of the
domain wall thickness with respect to the domain size.

Other messengers can come along the the domain walls from originating from the neighboring junction
and can be sources of approximate symmetries of our world.
Since we have anisotropic extra dimensions we might be able to see some of the
effects in high energy collider processes\mpp.

Another thing that the network of domain walls can affect is the cosmology of
early universe. Numerical simulations of 3+1 dimensional cosmology which admits
domain walls and their junctions show that a network of domain walls quickly
dominates the universe\vilenkin\rps . This can be a problem for cosmology if
the domain walls and their junctions reside in our observed universe. However,
if the domain walls are embedded in higher dimensional universe, and we are
living on the domain wall or on a junction, it is not going to be an immediate
problem. Senarios of cosmology can have phase transitons of the
domain wall configurations where the junctions dominate and where they do not.

While this paper was being typed the paper by S.M. Carroll, Hellerman and Trodden (hep-th/9911083) appeared
which discussed the BPS equations of domain wall junctions with local supersymmetry in a great detail.

\centerline{\bf Acknowledgements}

I have benefitted from useful conversations with V. Balasubramanian, C. Nunez,
R.~Gopakumar, A. Grant and K. Hori. 
This work is supported by KOSEF (981-0201-002-2) and by KRF(1998-015-D00073).

\listrefs
\bye